\documentstyle[preprint,prl,aps,amstex]{revtex}

\begin{document}

\title{The Equation of Motion of an Electron}

\author{R. F. O'Connell{\footnote{E-mail Address:
\emph{rfoc@@rouge.phys.lsu.edu}}}}

\address{Department of Physics
and Astronomy, Louisiana State University \\ Baton Rouge,
Louisiana, 70803-4001, USA}

\date{\today}

\maketitle

\begin{abstract}
The claim by Rohrlich that the Abraham-Lorentz-Dirac equation is not the
correct equation for a classical point charge is shown to be incorrect
and it is pointed out that the equation which he proposes is the
equation {\underline{derived}} by Ford and O'Connell for a charge with
structure.  The quantum-mechanical case is also discussed.
\end{abstract}

\newpage

\section{Introduction}

The equation of motion of a radiating electron is a subject of recurring
interest, and in particular, is discussed in the well-known book by
Jackson\cite{jackson98}.  For a classical point electron, Jackson used
the Larmor formula for the total power radiated by an accelerated charge,
in conjunction with conservation of energy arguments, to show that the
correct equation is the Abraham-Lorentz-Dirac (ALD) equation.  But, as
pointed out by Jackson, this equation has many well-known unsatisfactory
features.  This led Rohrlich\cite{rohrlich01} to question its validity.
However, the fact that an equation has unsatisfactory features does not
in itself mean that it is inconsistent with the assumptions underlying the
derivation.  In fact, Rohrlich does not point out anything wrong in the
derivation presented in Jackson\cite{jackson98}.  The fact is there is
nothing wrong!

Moreover, the equation which he proposes has various inconsistencies when
applied to a point electron, as was correctly pointed out by Ribaric and
Sustersic\cite{ribaric02} (a reply by Rohrlich\cite{rohrlich02} to this
paper was unconvincing) and by Baylis and Huschilt\cite{baylis02}.
Actually, these criticisms did not go far enough because they failed to
point out that the equation proposed by Rohrlich to describe a point
charge actually describes a charge with structure!  In fact the equation
analyzed by both \cite{ribaric02} and \cite{baylis02} is not actually the
relativistic Landau-Lifshitz equation but Rohrlich's approximate modification
of this equation.  As we point out below, the Landau-Lifshitz equation itself
was obtained by a series of approximations so there is no way of discerning
what they correspond to physically except to say that it is definitely not a
point particle.  This, coupled with the fact that both of these equations are
non-linear in the electromagnetic field tensor appears to me to be the
source of some of the inconsistencies pointed out by \cite{ribaric02} and
\cite{baylis02}.

Not only does Rohrlich's paper display inconsistencies but it
mischaracterizes previous work on this subject in that he claims that his
alternative equation to the ALD equation was obtained "-- by Ford and
O'Connell -- as an approximation --."  This is incorrect; Ford and
O'Connell\cite{ford91} obtained this exactly! Moreover, they showed
{\underline{exactly}} that it pertains to a particle with structure and
not to a point electron, as claimed by Rohrlich.  Moreover, in the
particular case of the point electron limit, they have, in essence,
obtained an exact derivation of the ALD equation for a classical point
charge starting with the universally-accepted Hamilton of non-relativistic
quantum electrodynamics(QED).  For these reasons and also because others
do not seem to have appreciated this point, we feel that it will be
instructive to summarize and extend the work of Ford and O'Connell, to point
out the nature of the problems with the Rohrlich paper and to outline how the
EXACT classical results, for both a point electron and an electron with
structure, may be obtained not only from the classical limit of the
quantum results but also from an ab initio classical derivation.

Moreover, since many important aspects of the problem arising from
quantum and fluctuation effects have been completely ignored in recent
discussions, we proceed in the following sections, in showing the
development from the classical case of a free electron to the quantum
case of an electron with structure, in a potential and subject to both
classical (temperature) fluctuations and quantum fluctuations.  Only
retardation has been neglected.  In particular, we attempt to point out clearly
results which have been derived in a precise and exact manner from carefully
stated initial assumptions \cite{jackson98,ford91}, in contrast to what I
refer to as "guesstimates" based on approximations which change the physical
nature of the problem in an unknown manner \cite{rohrlich01}.

\section{Nonrelativistic Classical Free Electron}

We first note, as points of reference, the classical Newtonian equation
of motion for a particle of mass $M$ under the action of an external
force $f(t)$

\begin{equation}
M\ddot{x}=f(t), \label{eme1}
\end{equation}
the corresponding AL equation for an electron of charge $e$

\begin{equation}
M\ddot{x} -M \tau_{e} \stackrel{\,\dots}{x} =f(t), \label{eme2}
\end{equation}
and the {\underline{exact}} equation derived by Ford and
O'Connell\cite{ford91} for a classical electron with a specific structure,
dipole interacting with the electromagnetic field,

\begin{equation}
M\ddot{x}=f(t)+\tau_{e}\dot{f}(t)=f_{eff}(t). \label{eme3}
\end{equation}
Here, a dot denotes differentiation with respect to $t$ and

\begin{equation}
\tau_{e}=\frac{2}{3}~\frac{e^{2}}{Mc^{3}}=6\times 10^{-24}s. \label{eme4}
\end{equation}

First of all, it should be emphasized that there is no conflict between
(\ref{eme1}) and (\ref{eme2}).  In particular, there are at least four
different but exact derivations for the case of a point electron all of
which lead to the same AL equation.  They are

\noindent (i) the well-known conventional derivation given in
Jackson\cite{jackson98}.

\noindent (ii) the point-electron classical limit of the exact
Ford-O'Connell results, given in (\ref{eme6}) below

\noindent (iii) the point-electron limit of a purely classical derivation
based on solutions of Maxwell's equations which was presented by Ford in
a comprehensive encyclopedia article\cite{ford931}.

\noindent (iv) the point-electron limit of a derivation in which the
electromagnetic field is treated as a classical stochastic field [see
discussion in the paragraph following (\ref{eme7}) below].

Three points are immediately clear:

\noindent (a) Because of the smallness of $\tau_{e}$, given in
(\ref{eme4}), the radiative reaction terms given in (\ref{eme2}) and

(\ref{eme3}) are both small compared to the other terms.

\noindent (b) There is no conflict between (\ref{eme2}) and (\ref{eme3})
because they describe different scenarios viz. a

point charge and a
charge with structure, respectively.

\noindent (c) Since from (\ref{eme1}) we have that

\begin{equation}
M\stackrel{\,\dots}{x}
=\dot{f}(t),~~~{\textnormal{to~order}}~\tau_{e}, \label{eme5}
\end{equation}
it follows that equations (\ref{eme2}) and (\ref{eme3}) only differ from
each other by terms of order $\tau_{e}^{2}$.  In other words, the
equations of motion for a point electron and an electron with structure
differ only by terms of order $\tau_{e}^{2}$ and higher order [the exact
difference being displayed in (\ref{eme16}) below].

Now, because (\ref{eme2}) existed since the 1903-1904 period and because
it was well-recognized that it had serious shortcomings
\cite{jackson98,oconnell96,ford98}, many attempts were made to modify
it.\cite{jackson98,eliezer48,landau75}.  In particular, one proposed
modification\cite{eliezer48} was, in effect, to add small
$\tau_{e}^{2}$ terms to (\ref{eme2}) to obtain (\ref{eme3}).

However, as we emphasized in \cite{ford91}, none of these
prior investigators derived (\ref{eme3}) from first principles.
Moreover, they surely knew that such additions to the AL equation of
motion clearly modified the basic assumptions underlying the AL
derivation as they were careful to avoid claiming that the resulting
equations represented the exact equation for a classical point electron.
Rohrlich, on the other hand, made this unwarranted claim.  In fact, his
"derivation" of (\ref{eme3}) from (\ref{eme2}) is, in essence, no
different than other ad hoc approaches; he uses the word "exact" liberally but
glosses over the fact that, in the discussion immediately following and
referring to his equation (\ref{eme3}), he states that "-- Since the latter is
a small term, the acceleration in it can be replaced by the approximate
expression from --."  In fact, he is using a "guesstimate" and he is
simply working to order $\tau_{e}$!  In essence, he is adding a term of order
$\tau^{2}_{e}$ without realizing that he is changing the physics of the problem
from a point electron model to a model with structure.

The key question to be answered is whether or not (\ref{eme3}) can be
derived exactly and what new physics does it incorporate.  The answer is
that this has already been done\cite{ford91}.  In \cite{ford91} we
presented an exact quantum mechanical result for the equation of motion
of the radiating electron from which we obtained (\ref{eme3}).  Our
starting point
\cite{ford85,ford88} was the universally accepted non-relativistic
Hamiltonian of QED describing an electron dipole interacting with a
radiation field and in an arbitrary potential but generalized to include
an arbitrary form factor
$f_{k}$ (Fourier transform of the electron charge distribution) and also
allowing for the presence of a time-dependent external field.  Next, we
used the Heisenberg equations of motion to obtain equations of motion for
the dynamical variables of the electron and of the radiation field in
terms of each other.  This enabled us to write down the equation of
motion of an electron with charge $e$ and bare mass $m$, dipole
interacting with the electromagnetic field and moving in a potential
$V(r)$, in the form of a generalized quantum Langevin
equation\cite{ford85,ford88}

\begin{equation}
m\ddot{x}(t)+\int^{t}_{-\infty}dt^{\prime}\mu(t-t^{\prime})\dot{x}
(t^{\prime})+V^{\prime}(x)=F(t)+f(t), \label{eme6}
\end{equation}
where $m$ is the bare mass, $x(t)$ is the coordinate operator, $F(t)$ is the
operator-valued random (fluctuating) force, $f(t)$ is the external force, $\mu
(t)$ is the memory function and where the dot and prime denote, respectively,
the derivative with respect to $t$ and $x$.  This is an {\underline{exact}}
result and explicit values are known for $\mu (t)$ and $F(t)$ in terms of
the parameters of the heat bath (in this case the radiation field).  If
fact, the (three-dimensional) random force is simply the free-field
operator \cite{ford88} i.e.

\begin{equation}
\vec{F}(t)=-e\bar{E}(t), \label{eme7}
\end{equation}
where $\bar{E}(t)$ is the electric field associated with the radiation
field.  In fact, such a term is crucial in order to satisfy the
fluctuation-dissipation theorem\cite{ford98}.

Equation (\ref{eme6}) is extremely general in that the choice of the
electron charge distribution has not been specified.  In particular,
specific choices for $\mu (t)$ lead to a variety of equations of motion,
in particular (\ref{eme2}) and (\ref{eme3}).  Furthermore, the
corresponding classical equation has exactly the same form as
(\ref{eme6}) but now the dynamical variables are all classical
quantities.  In fact, it is of interest to indicate how the classical
results can be obtained ab initio.  Completely analogous to the derivation
outlined above, we start with the classical Hamiltonian of
electrodynamics.  Next, we use Poisson bracket equations of motion and
obtain an equation of motion which is exactly the same form as
(\ref{eme6}) but now all the quantities are c-numbers.  It should be
emphasized that there are also fluctuations in the classical case (due to
temperature), whereas in the quantum case, there are both temperature and
quantum fluctuations.  Thus, it is clear that the quantum and classical
cases can be discussed in tandem.

Among other things, one advantage of the approach which we have adopted
is that it enables us to make use of the powerful techniques of
stochastic physics.  In particular, denoting the Fourier transform of
$\mu (t)$ by $\tilde{\mu}(\omega)$, it was shown \cite{ford88,ford91}
that it is analytic in the upper half of the $\omega$ plane and that it
has a positive real part.  This is equivalent to the demand
\cite{ford85,ford88} that all the poles of $\alpha (\omega)$, the
generalized susceptibility, must be in the lower half of the complex
plane (as otherwise the principle of causality is violated).  It
immediately follows \cite{ford91} that the AL point electron solution
does not fulfil these criteria and that we must consider an electron with
structure.

To proceed further, it is necessary to choose a particular form factor
(but, as outlined below, it can be shown that our results are correct to order
$\tau_{e}$ for all reasonable choices) and we chose an electron
form-factor
\cite{ford91}, $\Omega^{2}/(\Omega^{2}+\omega^{2})$, $\grave{a}$ la
Feynman \cite{feynman49}, with a shape convenient for calculation but
arbitrary in the sense that it depends on the choice of $\Omega$, a large
cut-off frequency.  Also, as is generally the case in quantum field
theory, mass renormalization is also required and we found that the
renormalized mass $M$ is given in terms of the bare mass $m$ by the
relation

\begin{equation}
M=m+2e^{2}\Omega /3c^{3}=m+\tau_{e}\Omega M. \label{eme8}
\end{equation}
Choosing $\Omega\longrightarrow \infty$ would correspond to the case of a
point electron.  But, as is clear from (\ref{eme8}),
$\Omega>\tau_{e}^{-1}=1.60\times 10^{23}s^{-1}$, would require a negative
rest mass and, based on the analyticity arguments given above, this is
precluded by physical considerations.  We conclude that

\begin{equation}
\Omega\leq\tau_{e}^{-1} \label{eme9}
\end{equation}
Thus, the choice $\Omega=\tau_{e}^{-1}$ corresponds to $m=0$ and
describes an electron with the smallest size consistent with causality.
The size is of the order of the classical electron radius \cite{oconnell93},
underlining the fact that the dipole interaction is more than adequate
to describe the basic physics.  [Nevertheless, it would be of interest to
calculate possible retardation modifications \cite{oconnell} to (\ref{eme6}),
a topic we hope to consider in the future.] This choice gives

\begin{equation}
f^{2}_{k}=\frac{1}{1+\omega^{2}\tau^{2}_{e}}, \label{eme10}
\end{equation}
with the result that the exact equation of motion reduces to

\begin{equation}
M\ddot{x}(t)+V^{\prime}_{eff}(x)=F_{eff}(t)+f_{eff}(t), \label{eme11}
\end{equation}
where

\begin{equation}
f_{eff}(t)\equiv f(t)+\tau_{e}\dot{f}(t), \label{eme12}
\end{equation}
and, similarly, for the other "effective" quantities.

In summary, based on the choice of form factor given by
(\ref{eme10}), (\ref{eme11}) is an {\underline{exact}} quantum mechanical
result for an electron with a minimum size (which turns out to be of the order
of the classical electron radius
\cite{oconnell93}) consistent with causality.  However, we will defer
discussion of the quantum aspects until Sec. 3.  We will now examine some
interesting limiting cases.

\noindent {\underline{(i)~~Classical Limit (with $V=0$)}} \\
\noindent The result has the same form as (\ref{eme11}) but with the
operator quantities replaced by c-number quantum mechanical averages and
with $V=0$.  Hence

\begin{equation}
M\ddot{x}=f_{eff}(t)+F_{eff}(t), \label{eme13}
\end{equation}
where, as before, $f(t)$, is the external force and the fluctuation force
$F(t)=-eE(t)$, where $E(t)$ is the classical electric field associated
with the radiation field.  Only in the case where one takes an ensemble
average so that

\begin{equation}
\langle E(t)\rangle =0 \label{eme14}
\end{equation}
does this result reduce to (\ref{eme3}).  On the other hand, in the
absence of an external force, we have

\begin{equation}
M\ddot{x}=F_{eff}(t),~~~{\textnormal{(if}}~f(t)=0). \label{eme15}
\end{equation}
This stochastic equation is intriguing, since it displays fluctuations
(the right hand side is a fluctuating force due to temperature effects
which, classically, give zero when the temperature $T=0$) with no
dissipative term on the left hand side.  Nevertheless, the
fluctuation-dissipation theorem still holds, as discussed in detail in
Ref. \cite{ford98}.

Now that we have established that (\ref{eme3}) is an exact result for the
classical equation of motion of an electron with a specified structure
and in the case where $V(x)=0$ and where an ensemble average has been
taken, we turn now to some further ramifications:

\noindent (ii)~~Equation (\ref{eme1}) may be re-written as an exact
equation of motion involving derivatives of $x$ (instead of the single
derivative of $f(t)$) but it requires an infinite number of terms in
powers of $\tau_{e}$ and derivatives of $x$ \cite{ford95}:

\begin{equation}
Mx^{(2)}+M\sum^{\infty}_{n=3}(-1)^{n}\tau_{e}^{n-2}x^{(n)}=f(t),
\label{eme16}
\end{equation}
where $x^{(n)}$ indicates the $n$ derivative of $x$ with respect to $t$.

\noindent (iii)~~If one again chooses the Feynman form factor but with
$\Omega <\tau_{e}^{-1}$ (as distinct from the value of $\Omega
=\tau_{e}^{-1}$ chosen above) then higher order terms in $\tau_{e}$
appear on the right-side of (\ref{eme3}), as discussed in \cite{ford91}.

\noindent (iv)~~The rate of radiation, $P(t)$ say, associated with the
equation of motion (\ref{eme3}) is no longer the Larmor rate but, as we
proved by two different methods \cite{ford912,ford913}, we now have

\begin{equation}
P(t)=\frac{\tau_{e}}{M}f^{2}(t)=\frac{2e^{2}}{3c^{3}}[f(t)/M]^{2}.
\label{eme17}
\end{equation}
In other words it differs from the Larmor result by terms of order
$\tau_{e}^{2}$.

We should emphasize once more that (\ref{eme11}) is an exact result
corresponding to the particular choice of form factor given by (\ref{eme10})
but now we will present a simple argument to show that (\ref{eme11}) is
actually correct, to a very good approximation, for all physically reasonable
choices of electron structure.  We proceed by considering another reasonable
choice of form factor and also, without loss of generality, we take $V(x)=0$.
Thus, for example, the choice

\begin{equation}
f^{2}_{k}=\frac{\Omega^{4}}{(\omega^{2}+\Omega^{2})^{2}}, \label{eme18}
\end{equation}
and $\Omega=2\tau^{-1}_{e}$, corresponding to a sharper cut-off, leads to

\begin{equation}
M\ddot{x}(t)=\left(1+\frac{\tau_{e}}{2}\frac{d}{dt}\right)^{2}\{F(t)+f(t)\}.
\label{eme19}
\end{equation}
Also, the typical heat bath correlation time, which is a measure of the
time-scale over which $F(\tau )$ changes on the average, is \cite{ford88}
$\hbar /(2\pi kT)=(1.1\times 10^{-12}/T)s^{-1}$, which is very large compared
to $\tau_{e}=6\times 10^{-24}s$.  Thus, it is an excellent approximation to
consider only lowest order in $\tau_{e}$, in which case we recover
(\ref{eme13}).  It is clear that, by a similar argument, all physically
reasonable choices of electron structure lead to (\ref{eme13}).

Thus, to lowest order in $\tau_{e}$ (an excellent approximation),
(\ref{eme13}), and more generally (\ref{eme11}), are model-independent results
for all choices of electron structure fulfilling reasonable physical criteria.

\section{Nonrelativistic Quantum Mechanical Equation}

The exact result for an arbitrary distribution of charge is given by
(\ref{eme6}) with all the dynamical variables to be regarded as operator
quantities.  For the particular choice of form factor given by
(\ref{eme10}), the resultant equation is (\ref{eme11}) i.e. it has the
same form as the classical equation.  Also, when $V=0$ and $f(t)=0$, the
resultant equation has the same form as (\ref{eme15}) but there is a
crucial difference in the quantum case viz. the right-side of
(\ref{eme15}) now contains quantum fluctuations as well as temperature
fluctuations.  Thus, even at $T=0$, there is a contribution from the
ubiquituous zero-point oscillations of the electromagnetic field (viz.
the same fluctuations that give use to the Lamb shift and the Casimir
effect)!  Furthermore, one may verify that such fluctuation terms are
essential to ensuring that the correct quantum-mechanical commutation
relations for the dynamical variables hold.

Finally, returning once more to our general quantum-mechanical result
(\ref{eme11}), it is instructive to consider the case of an electron in the
oscillator potential

\begin{equation}
V(x)=\frac{1}{2}Kx^{2}. \label{eme20}
\end{equation}
It then follows [see (\ref{eme12}) and the following line] that

\begin{eqnarray}
V_{eff}(x) &=& \frac{1}{2}K\left\{x^{2}+\tau_{e}\frac{d(x^{2})}{dt}\right\}
\nonumber \\
&=& \frac{1}{2}Kx^{2}+K\tau_{e}x\dot{x}. \label{eme21}
\end{eqnarray}
Thus, we obtain exactly the equation of motion.

\begin{equation}
M\ddot{x}+\zeta\dot{x}+Kx=-eE(t), \label{eme22}
\end{equation}
where $\zeta =K\tau_{e}$.  The presence of an Ohmic-type
dissipation term is now manifest.  In fact, this term is exactly of the
form universally used in practical applications of the AL equation to
radiation damping problems (where it is derived as a weak coupling
approximation to the dissipation term \cite{jackson98}).  It is also of
interest to note that $\zeta/\sqrt{KM}$ can be identified as a
dimensionless measure of the weak coupling (of the order of $10^{-7}$ for
optical frequencies) of relevance for studies in areas such as quantum
optics.

\section{Relativistic Equation of Motion}

Analogous to Dirac's extension of the AL equation to the relativistic
domain, we also carried out an analogous extension of (\ref{eme3}) with
the result \cite{ford93}

\begin{eqnarray}
Ma^{\mu} && =\frac{e}{c}F^{\mu} {_{\kappa}u^{\kappa}} \nonumber \\
&&~{}+\tau_{e}\frac{e}{c}\left(\frac{d}{d\tau}F^{\mu}
{_{\lambda}u^{\lambda}}-\frac{1}{c^{2}}u^{\mu}u^{\kappa}\frac{d}{d\tau}F_{
\kappa\lambda}u^{\lambda}\right), \label{eme23}
\end{eqnarray}
where

\begin{equation}
u^{\mu}=\frac{dx^{\mu}}{d\tau},~~a^{\mu}=\frac{du^{\mu}}{d\tau},
\label{eme24}
\end{equation}
with

\begin{eqnarray}
d\tau &=&
\frac{1}{c}\sqrt{g_{\kappa\lambda}dx^{\kappa}dx^{\lambda}}=\sqrt
{1-v^{2}/c^{2}dt} \nonumber \\
&=& \frac{1}{\gamma}dt. \label{eme25}
\end{eqnarray}
Also, the field tensor has the following form when expressed in terms of
the laboratory electric and magnetic fields,
\begin{equation}
F^{\mu v} =\left(
\begin{array}{rrrr}
0 				& -E_{x} & -E_{y} & -E_{z} \\
E_{x} & 	0 				& -B_{z} & B_{y} \\
E_{y} & B_{z} 	& 0 					& -B_{x} \\
E_{z} & -B_{y} & B_{x}  & 0
\end{array}
\right). \label{eme30}
\end{equation}
It is also of interest to note that
our equation of motion (\ref{eme3}) can be written in the three-vector form

\begin{equation}
M\frac{d\gamma
\vec{v}}{dt}=\vec{F}+\tau_{e}\left[\gamma\frac{d\vec{F}}{dt}-\frac{\gamma^{3}}
{c^{2}}\left(\frac
{d\vec{v}}{dt}\times (\vec{v}\times \vec{F})\right)\right], \label{eme31}
\end{equation}
where

\begin{equation}
\vec{F}=e(\vec{E}+\vec{v}\times \vec{B}) \label{eme32}
\end{equation}
is the Lorentz force.  In the quantum mechanical case, there will also be
relativistic fluctuation terms.

Our relativistic form (\ref{eme23}) followed from the non-relativistic result
(\ref{eme3}) by simply replacing $\ddot{x}$ by $a^{\mu}$ and $f(t)$ by

\begin{equation}
f^{\mu}=\frac{e}{c}F^{\mu} {_{\kappa}u^{\kappa}}, \label{eme33}
\end{equation}
where $F^{\mu}$ is the external electromagnetic field tensor.  Also
$\dot{f}(t)$ is replaced by

\begin{equation}
g^{\mu}=\dot{f}^{\mu}-\frac{1}{c^{2}}u^{\mu}u^{\kappa}\dot{f}_{\kappa},
\label{eme34}
\end{equation}
to ensure that

\begin{equation}
g^{\mu}u_{\mu}=0, \label{eme35}
\end{equation}
which is valid for any force four-vector.

Landau and Lifshitz also have written down a classical relativistic equation
\cite{landau75} but it can be classified as a "guesstimate" in that it is
derived from the Dirac relativistic result by a series of approximations.  In
other words, their starting point is based on the result for a point particle.
They then use a series of iterations (approximations) to express the result
directly in terms of the field tensor but, of course, these changes correspond
to a change in the underlying physics without any knowledge as to what these
changes are.  Thus their resultant equation has no discernable special
significance despite Rohrlich's claim \cite{rohrlich01}, and, of course, it
should be emphasized that Landau and Lifshitz never made such a claim. It is
interesting to note that the Landau-Lifshitz equation is non-linear in the
field tensor but whether or not non-linear terms will appear in the ultimate
theory is a matter of conjecture at this stage. Also, by construction, it is
purely classical (with $T=0$) and thus does not have fluctuation terms.

It clearly would be desirable to develop a relativistic theory ab initio but
incorporating electron structure into such a theory constitutes a formidable
task.  A start in that direction has been made by Johnson and Hu
\cite{johnson02} whose starting-point was the action describing the interaction
of a relativistic free particle interacting with a quantum scalar field (as
distinct from the quantum electrodynamic field).  Then, using stochastic
methods, these authors derived a relativistic Langevin equation leading to
causal solutions in the non-relativistic limit and they obtained results
consistent with our results.  Of course, as Johnson and Hu point out, some of
the results obtained are features of the assumption of a scalar field but they
certainly motivate their ultimate goal to extend the work to the relativistic
QED case.

\acknowledgments{The author is pleased to acknowledge a long-time
collaboration with G. W. Ford which led to the essential results presented
here.}

\end{document}